\documentclass[12pt]{article}
\usepackage{bm}
\usepackage{graphicx}

\title{QCD AGAINST BLACK HOLES OF STELLAR MASS $^*$}
 
\vspace{5mm}

\author{Ilya I. Royzen}
\date{}
\begin{document}
\maketitle \centerline{\em{P.N. Lebedev Physical Institute of RAS}}

\centerline{e-mail: $<royzen@lpi.ru>$}
\addtolength{\baselineskip}{6pt}
\vspace*{20mm}
\begin{abstract}

%\newpage%\vspace{10mm}

% \section{\bf GENERAL MOTIVATION} 
% \vspace{2cm}

In course of the consolidation of nucleon (neutron) spacing inside a compact star, two key 
factors are expected to come into play side by side: the lack of  
self-stabilization against shutting into black hole (BH) and forthcoming phase transition 
- color deconfinement and QCD-vacuum reconstruction - within the nuclear matter 
the star is composed of. These phenomena bring the star to evolve in the quite different 
(opposite) ways and should be taken into          
account at once, as the gravitational compression is considered. Under the 
above transition, which is expected to occur within any supermassive neutron star (NS), the 
hadronic-phase (HPh) vacuum - a coherent state of gluon- and 
chiral $q\bar q$-condensates - turns, first near the star center, into the ''empty''
(perturbation) subhadronic-phase (SHPh) one and, thus, pre-existing (very high)  
vacuum pressure falls there down rather abruptly; as a result, the ''cold'' star starts 
collapsing almost freely into the new vacuum. If the stellar mass is sufficiently 
large, then this implosion is shown to result in an enormous heating within the  star central 
domain (up to a
temperature about 100-200 MeV or, maybe, even higher), what makes the  pressure from within 
to grow up, predominantly due to degeneracy breaking and multiple $q\bar q$-pair production. 
Thus, a ''flaming wall'' could arise, which withstands the further collapsing and brings the star 
off the irrevocable shutting into BH. Instead, the star either forms a transient quasi-steady state 
(just the case of relatively low star mass) and, losing its mass, evolves gradually into the 
''normal'' steady NS, or is doomed for self-liquidation in full (at higher masses).

\end{abstract}

\vspace*{2.50cm}
----------------------------------------------------------------------------- 

$^*$ This is mainly the updated version of ref. [7] which is, in particular, free from some 
unnecessary assumptions.

\newpage

Two (incompatible) mechanisms underlying possible instability of a supermassive compact star are 
confronted below. Once it has arisen, they bring the star to evolve in absolutely alternative ways: 
the first (gravitationally-motivated) one, which is rather familiar, ''pushes'' it into a BH, 
whereas the second (QCD-motivated) one implies HPh $\to$ SHPh transition within the nuclear matter 
(it is considered here in more detail) and, {\it if activated in advance}, ''seeks'' to prevent it 
from approaching the BH horizon. In what follows, some indicative reasonings in 
favor of the BH-eliminating scenario are put forward.

\subsection{Phase transition in nuclear medium}

The main peculiarity of QCD is that, at non-extremal conditions (not very high
pressure and/or temperature),  the vacuum state is composed of the
quark-gluon condensate of high (negative) energy 
density, $\varepsilon^0_{vac} \simeq\,-5\,10^{-3}${GeV$^4$}, which shows itself up evidently in a 
number of elementary particle interactions \cite{Nov,Eid-Kur,V-Shif,Shif-Z}, and, hence, 
of the same modulo (but positive)  
pressure, $P^0_{vac} =\,-\varepsilon^0_{vac}$. One can also express this fact by the statement 
that the QCD Hamiltonian is {\it nearly} diagonalized with the following set of eigenstates:                                                            
vacuum condensate (the necessary entity) plus any possible configurations of nucleons and other 
hadrons which refer to a given baryon number $B$. The accent is made here on the word ''nearly'',
which emphasizes the negligibility of the residual interaction between the real particles and 
vacuum condensate under consideration of macroscopical (thermodynamical) processes at the 
''ordinary'' conditions, which are realized, actually, within any rarefied media - up to the 
within of the steady neutron stars (with masses $M_{NS}$ below the observed upper limit, 
$\overline{M_{NS}}\,\simeq\,2M_{\bigodot}$). To this extent, the vacuum condensate may be considered 
''a spectator'', which marks the zero-energy (and pressure) level. However, this idealization 
ceases to be physically relevant, when the gravitational compression makes the nuclear medium, 
first of all near the star center, sufficiently dense: then, two entities - the condensate and 
substance \footnote{For brevity, this term is addressed to everything, except of 
the vacuum condensate itself.} - start to affect each other considerably and destructively.
As a result, the vacuum condensate is no longer being a spectator, in the end it vanishes 
as well as the nucleon mass (the nucleons disintegrate into the massless quarks), and the 
eigenstates of QCD Hamiltonian become the proper configurations of colored particles - deconfined 
quarks and gluons.
In other words, the nuclear matter transforms into SHPh. This ''metamorphosis'' is necessarily to be 
taken into account as the evolution of a NS of ultraboundary mass ($M_{NS}\,\geq\,\overline{M_{NS}}$) 
is considered.

The key question is: whether some conditions exist which allow the NS nuclear matter to transform 
from HPh to SHPh without losing stability? In other words, is there a way of passing the 
interphase boundary (layer)
without undergoing  an enormous heating up? If yes, then no objections were, most probably, remained  
in the eyeshot against the statement that any sufficiently massive NS must finally shut into a BH 
\footnote{Actually, namely this logical sequence is meant when the OTO prediction of the ''imminent'' 
existence of stellar mass BH's is declared.}. However, according to the reasons given below, one has 
to say: No - the HPh $\to$ SHPh transition must show up certain unremovable explosion features 
\cite{R_2008,R_2009,IR}.

Physical motivation for this choice is, actually, quite simple - it is based on the crucial 
difference between EoSs for substance and vacuum: the former one is, evidently, 
$P\,\leq\,\varepsilon /3$, whereas the latter one is much harder, $P_{vac} =\,-\varepsilon_{vac}$. 
Since the NS periphery consists of a nuclear matter in (still relatively
rarefied) HPh, the total (vacuum + substance) pressure there is expected to be slightly higher than 
that inside the 
vacuum alone, $P^0_{vac} =\,|\varepsilon^0_{vac}|$ \footnote{By the way, within the framework of the 
Bag model,
namely the pressure $P^0_{vac}$ is responsible for making such the nucleons (and other baryons) 
as they are really.}.  Obviously, it cannot be lower near the (not collapsing) star 
center. Consequently, if the transition into SHPh (with {\it empty} vacuum) occurred there, then the
substance energy density (now it becomes the total one) is, wittingly, higher than 
$3|\varepsilon^0_{vac}|$. Meanwhile, near the outside of the phase transition boundary (where HPh 
is still retained!), the substance energy density, in any case, may not exceed that of the 
closely-packed nucleon (neutron) medium, because, 
otherwise, the particle wave functions would overlap so much that the individual nucleons loss their
identity and may no longer confine the quarks they were constructed from before
\footnote{That is tunneling (percolation), what may, probably, result effectively in color 
deconfinement somewhat earlier.}. That is why
$\varepsilon\,\leq\,\varepsilon_n\,\simeq\,|\varepsilon^0_{vac}|$ there, $\varepsilon_n$ being
the mean energy (mass) density within an isolated neutron.  Thus, the phase transition under 
consideration is definitely associated with, at least, tripling the substance energy density. The 
only reason for such a sharp enhancement seems to be very fast heating the medium which is followed
by degeneration breaking and multiple production of $q\bar q$-pairs and gluons.  

One can easily suggest the qualitative dynamics of this transformation: as the color charges  
are getting randomly unleashed, they start to violate
the long-range correlations in color field, which are responsible for vacuum condensate appearance; 
the condensate suffers of damages \footnote{One can say, figuratively, that the energy necessary for 
the condensate destruction is mainly taken at the price of vanishing of nucleon masses. This 
indicates
once more the deep interconnection between the vacuum condensate and nucleon mass in HPh.} 
and its pressure diminishes (the EoS of the entire substance-vacuum configuration softens); 
that is why the gravitational collapse is accelerating which
completes the condensate destruction and makes the nuclear substance to heat up strongly, 
initially near the star center \footnote{In this connection, one has to point out that thermal
neutrinos get essentially stuck at the relevant densities of nuclear matter and, therefore, the 
energy transport towards the star outside is an extremely slow process (a few hours {\it vs}
a typical hydrodynamic-time scale, which is, probably, of some fractions of a second).}, 
enforcing it to transform into {\it non-degenerate} state of SHPh, 
as a result the pressure of the substance there being increased again. Then, the collapse may stop 
and even turn into a turbulent swelling, unless the BH horizon
had time to emerge before in course of star compression \footnote{It is just the above-mentioned 
''competition to stay ahead'' between the two types of instabilities.}. In what follows, our goal 
is to show that, actually, there is not enough time for it to do. But firstly, some simple
estimates are made of how  the ultraboundary mass, $M_{NS}\,>\,\overline{M_{NS}}$ and corresponding 
radius $r$ of the central domain taken by SHPh are interrelated.

For brevity, in what follows, the SHPh-domain hot quark-qluon plasma (QGP) 
is referred a nearly perfect gas, which consists of 
the unremovable ''primordial'' quarks (carrying the net 
baryon-over-antibaryon surplus) as well as of the multiply produced qluons and
$q\bar q$-pairs, baryonic chemical potential $\mu_B$ thus tending to zero \footnote{The direct 
lattice MC simulations have shown \cite{Karsch} that, in this case, the thermodynamics 
of the real subhadronic medium mimics properties of such a gas within the accuracy of 20\%. Thus,   
the reasoning we put forward below keeps valid anyway.}. 
If the QGP domain is much less than the total star volume, then the obvious energy-conservation
equation reads:

\begin{equation}
-\,AG \frac{M_{NS}^2}{R^2_{NS}} dR_{NS}\,\simeq\,
4\pi \sigma_{QGP} \langle T^4 \rangle\,(1\,+\,
\frac{|\varepsilon^0_{vac}|\,-\,\varepsilon_n}{\sigma_{QGP} \langle T^4 \rangle}) r^2 dr,
\end{equation}

where on the left-hand side stands the work made by the gravitational field
($M_{NS}$ and $R_{NS}$ are the NS mass and its radius, respectively, and
the value of coefficient $A$ is 
confined in between of its non-relativistic and ultra-relativistic limits, 
$\frac{6}{7}\,\leq\,A\,\leq\,\frac{3}{2}$ \footnote{Below, we put $A\,=\,1$, 
since, in fact, the ultra-relativistic limit is rather inaccessible for the 
HPh-medium \cite{land}. For the same reason, the eq.(1) also neglects the role of particle 
pressure.}), whereas on the right-hand side stands the energy increase within 
the central QGP domain of a radius $r\,\ll\,R_{NS}$ and mean temperature $\langle T^4 \rangle$,   

$$\sigma_{QGP}\,=\,\frac{\pi^2}{30}[2\,\times\,8\,+
2\,\times\,3\,\times\,2\,\times\,(2\,\div\,3)\,\times\,\frac{7}{8} +(12 \div 16)]$$

being the weight factor of $(2\,\div\,3)$-flavor QGP (8 gluons of spin 1 and (3 + $\bar 3$) 
colored quarks of spin 1/2, plus photon and lepton contribution (the last item)) \footnote{Some 
numerical luft results from the fact that the temperature (see below) and $s\bar s$- and 
$\mu^+ \mu^-$-pair masses may occur of the same order, thus the relevant freedom degrees being only
half-alive. What is, principally, more essential is that, unlike                                            
the QGP-fireball produced in heavy ion collisions, the photons and leptons participate
now on an equal with quarks and gluons in the establishing of an equilibrium state.}. 
In the accordance with what was mentioned above, the inequality

\begin{equation}
\sigma_{QGP} \langle T^4 \rangle\,\geq\,3 |\varepsilon^0_{vac}| 
\end{equation}

should take place, wherefrom one can find a lower limit for the mean QGP temperature:

\begin{equation}
\langle T^4 \rangle ^{1/4}\,\geq\,160 MeV
\end{equation}

This temperature is, at least, about 20-30 times higher than typical temperatures of supernova  
explosions and, therefore, of the neutron medium outside the SHPh. Thus, the hydrodynamic (fast
process) balance can be achieved (if it ever possible) only at the cost of   
an enormous thermal disbalance. It is worth mentioning here an encouraging correlation between the 
above estimate and result of well known lattice MC simulation \cite{Karsch}, which indicates, at 
$\mu_B=\,0$, 
the HPh $\to$ SHPh  crossover within the temperature range 140\,MeV\,$\leq\,T\,\leq$\,200\,MeV.
              
Being combined, the relations (2) and  $\varepsilon_n\,\simeq\,|\varepsilon^0_{vac}|$ suggest 
that one can neglect the second term in the brackets on the right-hand side of eq.(1)
\footnote{Anyway, this level of accuracy is suitable in the relevant estimates.}. Thus, integrating 
eq.(1), one obtains

\begin{equation}
G \frac{M_{NS}^2}{R_{NS}}\,=\,\simeq\,
\frac{4\pi}{3} \sigma_{QGP} \langle T^4 \rangle r^3\,+\,C,
\end{equation}

where $C$ is defined by $\overline{M_{NS}}$ - the value of mass upper limit 
for the {\it really} 
steady (''cold'' everywhere, i.e., $r$ = 0) NSs: $C\,\simeq\,(0.5\,\div\,1)\,M_{\bigodot}$
for $\overline{M_{NS}}\,\simeq\,(1.5\,\div\,2.5) M_{\bigodot}$ 
and $R_{NS}\,\simeq\,(8\,\div\,10)$ km, respectively \footnote{The current observations favor 
definitely the lower of these estimates.}. Of course, the
transient ''quasi-steady'' heterogenic mode of a high-mass NS nuclear medium, described by eq.(4),
could be thought as physically realizable, only if $r\,\ll\,R_{NS}$, thus HPh $\to$ SHPh transition 
being not too violent (i.e., if the temperature profile is not too sharp \cite{land1}). Then, it 
seems 
sensible to imagine a rather quiet combustion within the supermassive NS which does not turn into
detonation in full. This process is, undoubtedly, 
accompanied by some eruptions of star substance and/or gamma bursts, both being the more powerful
the more noticeable is the difference ($M_{NS} -\,\overline{M_{NS}}$); this ''volcano activity''
results in a diminishing of star mass and is expired as $M_{NS}$ approaches $\overline{M_{NS}}$. 
At still larger  initial values of $M_{NS}$, the eq.(4) asks {\it formally} for $r\,\simeq\,R$, but 
it means nothing else than the fact that the very approximation adopted ceases to be admissible. 
Instead, one can 
reasonably expect that no room remains now for the achievement of a transient hydrodynamic balance 
and more or less quiet evolution: in this case, one can expect the development of 
powerful shock waves, which should forward NS towards the 
catastrophic self-destruction \footnote{From the more general point of view, the variety of
possible ways of stopping the star collapse demonstrates nothing else than that there are different 
ways of symmetry (in this context - of the chiral one) breaking along with medium cooling and 
getting more rarefied: the no-order-parameter 
SHPh turns into the HPh, which shows up clearly an order parameter - for it can be chosen, say,
the inverse radius of color confinement.}.

\subsection{BHs of the lowest mass, NSs of the highest mass and large  
            band gap in between. No way for compact star $\to$ BH evolution.}

At the same time, the horizon of a radius $R_{BH}$ is, obviously, emerged at the condition 
$\frac{2GM_{BH}}{R_{BH}}$ = 1, or, what is the same,

\begin{equation}
R_{BH}\,=\,\lbrack\frac{3}{8\pi 
G \langle\varepsilon_{BH} \rangle }\rbrack^{1/2,}
\end{equation}

where $M_{BH}$ and $\langle\varepsilon_{BH} \rangle$ are the BH mass and its mean 
energy density, respectively. For getting the lower estimate of $R_{BH}$ 
(in the context of compact star collapse), one has to take into account that 
$\langle\varepsilon_{BH} \rangle\,\leq\,\varepsilon_n\,\simeq\,|\varepsilon^0_{vac}|$, since,  
otherwise, the phase transition instability followed by the aforementioned consequences 
is expected to activate before. Thus, one obtains for $R_g$ = min$R_{BH}$ and $M_g$ = min$M_{BH}$

\vspace*{3mm}
\centerline{$R_g\,\geq\,12$ km or $M_g\,\geq\,4\,M_{\bigodot}$}  
\vspace*{3mm}

This is, actually, an underestimation (most probably, a considerable one) because the star 
interior density is higher than the 
peripheral  one, and, therefore, the hot SHPh matter starts forming there even earlier. Thus, 
the NSs of highest mass, ($M_{NS}\,\simeq\,\overline{M_{NS}}\,\leq\,2\,M_{\bigodot}$), which are 
observed  so far (and had a perspective to turn into BH), and the hypothetical BHs of lowest 
mass predicted by GR are separated by a very significant gap.
What kind of star organization could set up in between? If the NS mass still were imagined to access 
$4\,M_{\bigodot}$, then eq.(4) tells immediately that $r\,\simeq\,R_{NS}$, what means 
nothing else than, in fact, no sensible solutions exist at all.   
The only reasonable interpretation of this fact is, seemingly, that any collapsing star of a mass 
$M$, which considerably exceeds $2\,M_{\bigodot}$, 
is doomed for the complete destruction just after (hydrodynamic time scale!) the nucleon 
packing becomes sufficiently 
compact. Thus, a large (semi-phenomenologically/semi-theoretically motivated) band gap,  
$\Delta\,\geq\,2\,M_{\bigodot}$,
between the allowed NS (even transient one) and BH masses blocks the way for compact star $\to$ BH 
evolution.

We put aside everything that relates to the formation of less compact BH of substantially larger 
$M_{BH}$ and $R_{BH}$ 
(both are $\sim\,\langle\varepsilon_{BH} \rangle^{-1/2}$),  
since, in this case, a more detailed information on star dynamics should be involved. What can be 
said from the general considerations, is that shutting to BH of such a type is, seemingly,  
an event of even lower probability. Essentially, the matter is that the conditions, which make 
horizon to emerge, are linked to the $global$ features of the star nuclear 
medium (the values of $M/R$ and $averaged$ energy density 
$\langle\varepsilon \rangle$), whereas the HPh $\longrightarrow$ SHPh  
transition instability is linked directly to the $local$ values of 
$\varepsilon$, which, undoubtedly, increase towards the 
star center. Since the relevant EoS is rather soft (the medium is non-relativistic and thus
$P\,\ll\,\varepsilon /3$), one can expect this increase to be sufficiently steep
for making the proactive development of HPh $\longrightarrow$ SHPh transition instability near the 
star center.

%The same argument gives, all the more, ''obvious preference'' 
%to the proactive development of HPh $\longrightarrow$ SHPh instability in case 
%of some density fluctuations within the star body. 
%Thus, this instability is anyway expected to start developing at lower values 
%of $\langle\varepsilon \rangle$.  

It is worth also 
mentioning, in this connection, that some factors unaccounted above - unavoidable energy density
fluctuations, expected star rotation and its
non-sphericity and, especially, binary-star configuration -
should, obviously, result in diminishing the margin of star 
stability, thus making the 
above arguments against attainability of BH-horizon even more defensible.

\section{Conclusion}
%\vspace*{1cm}
The QCD-induced mechanism of NS instability is discussed which is incompatible with the 
gravitational one. The NSs of 
highest masses are proven to be in face of instability associated with 
QCD-vacuum transformation under HPh $\longrightarrow$ SHPh transition, which 
could manifest itself, in particular, through the softening of EoS towards the 
star center. This instability seems to develop before BH horizon emerges somewhere
within the star body, what makes rather improbable the very 
accessibility of a BH configuration at the end of collapsing star evolution.\\

Since the temperature of substance in SHPh formed near the star center is more than one order 
higher than that of the supernova explosions, the relevant energy release could be up to several 
orders higher. That is why one cannot rule out that some poorly understood observation  
data on very distant (''young'') and most powerful GRB's - 
like GRB 090423 \cite{krimm}, GRB 080916C \cite{abdo}, GRB 080319B  
(''naked eye'') \cite{bloom}, Sw 1644+57 \cite{CERN}, etc. - (which are associated sometimes with      
insatiable ''eating up'' the stellar substance by situated (supposedly) nearby BH) are linked,
actually, with the above phase instability within the neutron stars themselves.

\end{document}